\documentclass[twocolumn,aps,prl,epsfig,graphics]{revtex4}
\usepackage{graphicx}
\usepackage{amsmath}
\begin{document}

\title{Schr\"{o}dinger Cat States of a Nanomechanical Resonator}
\author{L. Tian$^{1,2,3}$}
\affiliation{$^{1}$ Institute for Theoretical Physics, University of 
Innsbruck, 6020
Innsbruck, Austria \\
$^{2}$ Institut f\"ur Theoretische Festk\"orperphysik, Universit\"at
Karlsruhe, D-76128 Karlsruhe, Germany \\
$^{3}$ Institute for Quantum Optics and Quantum Information of the Austrian
Academy of Sciences, 6020 Innsbruck, Austria }
\date{\today }

\begin{abstract}
We present a scheme of generating large-amplitude Schr\"{o}dinger cat states 
and entanglement in a coupled system of nanomechanical resonator and single Cooper pair 
box (SCPB), without being limited by the magnitude of the coupling. It 
is shown that the entanglement between the resonator and the SCPB can be 
detected by a spectroscopic method.
\end{abstract}
\maketitle

The fabrication and probing of ultra-small nanomechanical resonators with  
secular frequencies of $\mathrm{GHz}$ and quality factors approaching $10^5$ 
have been achieved in recent experiments\cite{ResonatorSET}. These resonators 
are promising systems for demonstrating the quantum mechanical nature of the 
mechanical degrees of freedom\cite{NanoResonator}. Potential applications 
include detection of weak forces, precision measurement, and quantum 
information processing\cite{forcedetection,Munro_measurement,QIP_resonator}. 
One crucial step in studying the nanomechanical resonators will be the 
quantum engineering and the detection of the mechanical modes. This can be  
achieved by connecting the resonators with solid-state electronic 
devices\cite{ResonatorSET,resonator_SCPB1,resonator_SCPB2}, for example 
coupling a resonator with a single electron transistor (SET) via 
electrostatic interaction. The SET measures the flexural oscillation of the 
resonator with an accuracy approaching the quantum limit\cite{ResonatorSET}.  
Cooling of a resonator to its ground state has been proposed by quantum 
feedback control via a SET\cite{feedback_cooling} and by side band cooling 
via a quantum dot\cite{wilsonrae_cooling} or a SCPB\cite{Resonator_cooling}.

The resonator modes can be treated as underdamped harmonic oscillators with 
the damping described by the finite quality factor $Q$. Connecting a 
resonator with a quantum two level system forms a spin-oscillator model which  
has been intensively studied in quantum optics, especially in ion trap 
quantum computing\cite{IonTrap_Rev}. Hence the techniques of manipulating the 
motional state of a trapped ion by laser control of its internal mode can be 
applied to studying the nanomechanical resonators\cite{IonTrap_Rev}. In 
Ref.~\cite{resonator_SCPB1,resonator_SCPB2}, the capacitive coupling between 
a nanomechanical resonator and a SCPB was studied, where the SCPB can be 
treated as a two level system -- the superconducting charge qubit -- by 
adjusting the parameters and the gate voltage\cite{charge_qubit_mss}. It was 
shown that entanglement between the resonator and the qubit can be generated 
and be detected by interferometry when the coupling is stronger than the 
energy of the resonator. In this paper, we show that large-amplitude 
Schr\"{o}dinger cat states and entanglement can be generated in the coupled 
resonator and SCPB system by parametric pumping of the SCPB when the 
magnitude of coupling is weak due to the geometry of the charge island and 
the distance between the charge island and the 
resonator\cite{Resonator_cooling}. Given the large amplitude of the generated 
cat states, the entanglement between the resonator and qubit can be observed 
by a spectroscopic measurement which selectively flips the charge qubit 
depending on the state of the resonator. When the scheme is generalized to 
two or more nanomechanical resonators, it generates maximal entanglement 
between these resonators, which is a key element in continuous variable 
quantum computing\cite{qc_braunstein}. In precision measurement, the cat 
state of $N$ resonators can increase the sensitivity to weak forces by a 
factor of $\sqrt{N}$\cite{Munro_measurement}. The effect of environmental 
fluctuations including the mechanical noise and the charge noise around the 
charge island is analyzed. Furthermore, as this scheme involves the generic 
system of one spin and one oscillator, it can be tailored for other 
applications such as single spin detection by a 
cantilever\cite{single_spin_detection}.

\begin{figure}[tbh]
\includegraphics[width=7.5cm,clip]{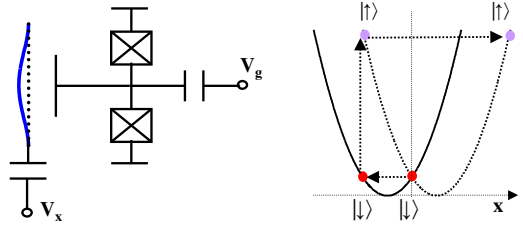}
\caption{Left: a resonator couples with a SCPB. Right: the trajectory of the 
resonator starts from the origin $x=0$ (thin dotted line) and the state $|\downarrow\rangle$ 
of the charge qubit; the qubit is flipped every half period of the resonator 
$\pi/\omega_{0}$.} \label{figure1}
\end{figure}
The coupled system of a nanomechanical resonator and a SCPB is shown in 
Fig.~\ref{figure1}, with the resonator undergoing flexural vibration. The 
flexural mode is described by the Hamiltonian $H_{m}=\hbar\omega_{0} 
\hat{a}^{\dag }\hat{a}$ where $\omega _{0}$ is the frequency of the mode and 
$\hat{a}^{\dagger }$ ($\hat{a}$) is the raising (lowering) operator of the mode. 
The resonator is biased at a voltage $V_{x}(t)$ and couples to the SCPB 
through a capacitance $C_{x}(\hat{x})=C_{x0}(1+\hat{x}/d_{0})$ where 
$C_{x0}$ is the static capacitance, $d_{0}$ the static distance between the
two, and $\hat{x}=\delta x_{0}(\hat{a}+\hat{a}^{\dagger })$ the displacement 
of the flexural mode with $\delta x_{0}=\sqrt{\hbar/2m\omega_{0}}$ the quantum width of the 
resonator. The SCPB is a superconducting island connected with Josephson junctions and 
is controlled by the gate voltage $V_{g}(t)$ through the gate capacitance 
$C_{g}$. When $C_{g}V_{g}+C_{x}V_{x} = (2n+1) e+2e \delta n$ with $n$ integer 
and $\delta n\ll 1$, the SCPB can be treated as an effective quantum two 
level system -- the superconducting charge qubit\cite{charge_qubit_mss} -- 
described by the Hamiltonian $H_{q}=4E_{c}\delta n 
\sigma_{z}+\frac{E_{J}(t)}{2} \sigma _{x}$ with $E_{c}=e^2/2C_{\Sigma}$ the 
charging energy, $C_\Sigma$ the total capacitance connected with the charge 
island, and $E_{J}(t)$ the Josephson energy with a small modulation around 
the static Josephson energy $E_{J0}$. Here $\sigma _{x,z}$ are Pauli 
operators for the two level system. To the lowest order, the coupling between 
the resonator and the SCPB is $-\lambda \left(t\right)\left (\hat{a}+\hat{a} 
^{\dag }\right) \sigma _{z}$ with $\lambda (t)=4E_{c} (V_{x}(t)C_{x0}/2e) 
(\delta x_{0}/d_{0})$, and its magnitude can be limited by $C_{x0}$ and the small ratio of 
$\delta x_{0}/d_{0}$. In our scheme, the bias of the resonator is 
$V_{x}(t)=V_{x0} \cos \left(\omega_{ac}t\right)$, an ac voltage with amplitude 
$V_{x0}$ and frequency $\omega_{ac}$; the gate voltage is $V_{g}(t) = 
V_{dc}+V_{g0}\cos\left(\omega_{ac}t\right)$, including a dc voltage $V_{dc}$ 
and an ac voltage with amplitude $V_{g0}$ and frequency $\omega_{ac}$. We let 
$C_{g}V_{dc}=(2n+1)e$ so that the charge qubit operates at the degenerate 
point\cite{optimal_point}. In experiments, it was shown that the decoherence 
time of the charge qubit at the degenerate point can be as long as 
microseconds\cite{optimal_point}. Here $\omega_{ac}=E_{J0}/\hbar$ is in 
resonance with the energy of the charge qubit at the degenerate point. In the 
following we study the system in the rotating frame of $\frac{E_{J0}}{2} 
\sigma_{x}$. In this frame the Hamiltonian after the rotating wave 
approximation can be derived as
\begin{equation}
H_{rot}=\hbar \omega _{0}\hat{a}^{\dag }\hat{a}-\frac{\lambda _{0}}{2}
\left( \hat{a}+\hat{a}^{\dag }\right) \sigma _{z}-\frac{\epsilon _{z}}{2}
\sigma _{z}+\frac{\epsilon _{\perp }\left( t\right) }{2}\sigma _{x}
\label{Htr}
\end{equation}
where $\lambda _{0}=4E_{c} ( V_{x0}C_{x0} /2e )(\delta x_{0}/d_{0} )$ is the 
coupling in the rotating frame, $\epsilon _{z}=8E_{c} 
(C_{g}V_{g0}+C_{x0}V_{x0})/2e \ll E_{c}$, and $\epsilon _{\perp }=E_{J}(t)-E_{J0}$ is 
the small modulation of the Josephson energy. Here, the dynamics of the 
resonator is that of a shifted harmonic oscillator with the Hamiltonian 
$DH_{m}D^{\dagger }$ with the displacement operator
\begin{equation*} D=\exp \left( -\left( \hat{a}-\hat{a}^{\dag }\right) 
\frac{\lambda
_{0}\sigma _{z}}{2\hbar \omega _{0}}\right).
\end{equation*}
Typical parameters\cite{Resonator_cooling} are $E_{J0} \approx 10\,\mathrm{GHz}$, 
$E_{c}\approx 50\,\mathrm{GHz}$, $C_{x0}\approx 20\,\mathrm{aF}$, and  $\omega 
_{0}\approx 100\,\mathrm{MHz}$. With $V_{x0}\approx 1\,\mathrm{V}$, the coupling is 
$\lambda_{0} \approx 20\,\mathrm{MHz}$. 

Below we show that by pumping the charge qubit with stroboscopic pulses, 
Schr\"{o}dinger cat states with large amplitude can be generated in this 
system. In an ideal situation, we consider $\delta$-function pulse sequence
\begin{equation}
\epsilon _{\perp }\left( t\right) =\pi \delta \left( n\tau _{0}\right ), 
\quad \mathrm{with}\,n\geq 1,\mathrm{\ \ }\mathrm{integer} \label{ep_perp}
\end{equation} 
where each pulse is a transformation $-i\sigma _{x}$ that flips the charge 
qubit after every half period of the resonator mode $\tau _{0}=\pi/\omega 
_{0}$. Here $\epsilon_{z}=0$. The $\delta $-function approximation is 
valid when $\epsilon _{\perp }\gg \hbar \omega _{0},\lambda_{0}$. Let 
$U_{1}=e^{-iH_{rot}\tau _{0}}|_{\epsilon_{\perp}=0}$ be the free evolution 
of the system between the pulses: $U_{1}=D e^{-i\pi \hat{a}^{\dag }\hat{a} } 
D^{\dag} $. At times $n\tau_{0}$ after the $n$th pulse, the unitary transformation on the system 
is 
$U\left(n\tau _{0}\right) = \left(-i\sigma_{x}U_{1}\right) ^{n}$. With 
the relations: $\sigma _{x}D\sigma _{x}=D^{\dagger }$ and $e^{ i\pi 
\hat{a}^{\dag }\hat{a}} D e^{-i\pi \hat{a}^{\dag }\hat{a}} =D^{\dagger}$, 
we derive 
\begin{equation}
\displaystyle U\left( n\tau _{0}\right) =\{ \begin{array}{ll}
(D^{\dagger })^{2n}, & n\in even \\ [2mm] \sigma_{x} e^{-i\pi \hat{a}^{\dag 
}\hat{a}} (D^{\dag} )^{2n}, & n\in odd
\end{array}\label{Ut}
\end{equation}
where the overall phase factors are omitted. This transformation generates 
in the wave function of the resonator a displacement of $\Delta x=-\delta 
x_{0}\left( 2n\lambda _{0}\sigma _{z}/\hbar \omega_{0} \right)$ when $n$ is 
even, and an opposite displace when $n$ is odd. With the initial state 
$|\psi _{0}\rangle =(c_{0}|\uparrow \rangle +c_{1} |\downarrow \rangle )\int 
dx|x\rangle \varphi (x)$,  where the states $|\uparrow,\downarrow \rangle $ 
are eigenstates of the charge qubit in the $\sigma _{z}$ basis and $\varphi 
(x)$ is the wave function of the resonator, after even number of flips $n$, 
the state is
\begin{equation}  \label{psi}
c_{0}\left| \uparrow \right\rangle \int dx\left| x\right\rangle \varphi
( x+\Delta x) + c_{1}\left| \downarrow \right\rangle \int dx\left 
| x\right\rangle \varphi ( x-\Delta x)
\end{equation}
where the state of the resonator is shifted according to the state of the 
charge qubit.

Assume an initial state of $|\psi_{0}\rangle = \frac{1}{\sqrt{2}} 
(|\uparrow\rangle +|\downarrow\rangle) |0\rangle$ where $|0\rangle $ is the 
ground state of the resonator and $\lambda_{0}=0$ at $t<0$. 
Following Eq.~(\ref{Htr}), after even number of pulses $n$, the state
$\frac{1}{\sqrt{2}}(|\uparrow~\rangle |-2n\alpha _{0}\rangle + |\downarrow 
~\rangle |2n\alpha_{0}\rangle )$ is generated where $\alpha_0 = \lambda_{0} 
/2\hbar\omega_{0}$ is the dimensionless coupling. The state 
$|\alpha\rangle$ denotes a coherent state of the resonator mode with an
amplitude $\alpha$. When $2n \alpha_{0} \gg 1$, 
maximal entanglement is generated between the resonator 
and the charge qubit. An intuitive way to describe the process is to 
consider a classical particle with two spin components in an harmonic 
potential, where the potential is shifted from the origin to the left 
(right) at spin down (up) when the coupling is on. The spin is subject to 
kicks every half period of the oscillator, as shown in Fig.~\ref{figure1}. 
At $t<0$, the oscillator is at the origin in its ground state with no 
coupling. At $t>0$, the coupling is turned on and the particle starts 
oscillating with opposite displacements for the two spins. Each kick 
maximally increases the energy of the particle and generates coherent 
states with large amplitude. Writing the generated state in the 
$|\pm\rangle$ basis, we have $\frac{1}{2}|+ \rangle (|-2n\alpha_{0}\rangle 
+|2n\alpha _{0}\rangle ) + \frac{1}{2}|- \rangle (|-2n\alpha_{0}\rangle 
-|2n\alpha _{0}\rangle )$. A measurement on the $\sigma_{x}$ operator of the 
charge qubit with the scheme in \cite{optimal_point} projects the resonator 
to the state $\frac{1}{\sqrt{2}}(|-2n\alpha _{0}\rangle \pm|2n\alpha _{0} 
\rangle )$ corresponding to the measured $\sigma_x$ value $+$ or $-$ 
respectively. Note that with the Hamiltonian in Eq.~(\ref{Htr}), 
entanglement can be generated at the degenerate states $|\uparrow\rangle |\alpha 
_{0}\rangle $ and $|\downarrow\rangle |-\alpha _{0}\rangle $ without the pumping 
process\cite{resonator_SCPB1,resonator_SCPB2}. However, with $\lambda 
_{0}\ll \hbar \omega _{0}$, the coupling only slightly shifts the resonator 
state: $\alpha _{0}\ll 1$ and the resonator is only weakly entangled with the 
charge qubit. With the pumping process, a shift of the wave function much 
larger than $\alpha_{0}$ can be achieved.

This scheme can be generalized to multiple resonators and (or) charge qubits to 
generate entanglement between the resonator modes. When two resonators 
couple with one charge qubit with the coupling $\sum \frac{\lambda_{0i}}{2} 
(\hat{a}_{1}+\hat{a}_{i}^{\dag})\sigma_{z}$, the state $\frac{1}{\sqrt{2}} 
(|-\alpha_1\rangle_{1}|-\alpha_2\rangle_{2}\pm 
|\alpha_1\rangle_{1}|\alpha_2\rangle_{2} )$ can be 
generated, where the index $i=1,2$ labels the two resonators and $\alpha_{i} 
= n\lambda_{i}/\hbar\omega_{0}$. Such states are maximally entangled states 
of the resonator modes\cite{kraus_pra} and are crucial elements in 
continuous variable quantum computing\cite{qc_braunstein}.

The entanglement and coherence between the resonator and the charge qubit 
can be detected by a spectroscopic method for resonator states of large 
amplitude with $2n\alpha_0 \gg 1$. During the detection, choose a static 
bias $\epsilon_{z} > 0$ of the charge qubit and modulate the Josephson 
energy by $\epsilon _{d}\cos \left( \omega _{d}t\right) \sigma _{x}$ 
with a frequency $\omega_{d}$ for a duration of $\pi/\omega_{d}$. Here 
instead of the $\delta$-function pulses in Eq.~\ref{ep_perp}, 
$\epsilon_{d}$ has the same order of magnitude as $8n\alpha_{0} 
\lambda_{0}$ and $\epsilon_{z}$, while $\epsilon_{d} \gg \omega_{0}$ and 
the resonator can be treated as static during the detection. This 
condition is crucial in realizing the detection process. The effective 
Hamiltonian of the SCPB is then:
\begin{equation}
H_{q}^{\left| \pm 2n\alpha _{0}\right\rangle}=-\frac{\epsilon
_{z}\pm 4n\alpha _{0}\lambda _{0}}{2}\sigma _{z}+\epsilon _{d}\cos \left(  
\omega _{d}t\right) \sigma _{x}  \label{Heff}
\end{equation}
for the resonator states of $\left| \pm 2n\alpha _{0}\right\rangle $ 
respectively. Hence the energy splitting of the charge qubit depends on 
the state of the resonator with a splitting $E_{+}=\epsilon_{z}+4n\alpha 
_{0}\lambda _{0}$ for the resonator state $\left| 2n\alpha 
_{0}\right\rangle $ and $E_{-}=\epsilon_{z}-4n\alpha _{0}\lambda _{0}$ for 
the resonator state $\left| -2n\alpha_{0}\right\rangle $. We choose the 
pulse frequency to be $\hbar\omega_{d}= \epsilon _{z}-4n\alpha _{0}\lambda 
_{0}$, in resonance with the charge qubit in $H_{q}^{ \left| -2n\alpha_{0} 
\right\rangle}$. This pulse is then followed by a $\delta$-function 
$\pi/2$ pulse that transforms $|\uparrow,\downarrow \rangle $ to 
$|+,-\rangle $. Applying the pulses to the state $\frac{1}{\sqrt{2}} 
(|~\uparrow~\rangle |~-2~n~\alpha _{0}~\rangle + |~\downarrow~\rangle |~2n~\alpha 
_{0}~\rangle )$, the final state is
\begin{equation}
-\frac{i}{\sqrt{2}}|-\rangle |-2n \alpha_{0}\rangle 
+\frac{1}{\sqrt{2}}\left( c_{\downarrow }|-\rangle +c_{\uparrow }|+\rangle 
\right) |2n \alpha_{0}\rangle  \label{psi_tau}
\end{equation}
with $c_{\uparrow }=-i\sin \left(\pi\bar{\epsilon}_{d}/2\epsilon_{d}\right)   
\left( \epsilon _{d}/\bar{\epsilon}_{d}\right)$ and
\begin{equation*}
c_{\downarrow }=\cos \left( \frac{\pi \bar{\epsilon}_{d}}{2\epsilon
_{d}} \right) -i\sin \left( \frac{\pi
\bar{\epsilon}_{d}}{2\epsilon _{d}}\right)
\frac{8n\alpha_{0}\lambda _{0}}{\bar{\epsilon}_{d}}
\end{equation*}
where $\bar{\epsilon}_{d}=\sqrt{\epsilon _{d}^{2}+\left (8n\alpha_{0} 
\lambda _{0}\right) ^{2}}$. For the resonator state $|2n\alpha_{0} 
\rangle$, the off resonance $8n\alpha _{0} \lambda _{0}$ between 
$\omega_{d}$ and $E_{+}$ prevents the charge qubit from flipping. By 
adjusting the bias $\epsilon_{z}$ and the amplitude $\epsilon_{d}$, we 
can find a regime where $|c_\downarrow|\approx 1$. Note the states 
$|\pm\rangle$ are in the rotating frame and in the lab frame the 
$\sigma_{x}$ eigenstates are $|\pm\rangle_{s}=e^{\pm iE_J t/\hbar} 
|\pm\rangle$. A measurement on the $\sigma_{x}$ operator of the 
charge qubit as in \cite{optimal_point} obtains the probabilities of the 
states $|\pm \rangle $: $p_{-}=(1+|c_{\downarrow }|^{2})/2$ and 
$p_{+}=|c_{\uparrow }|^{2}/2$ respectively. As a first step, the 
correlation between the resonator and charge qubit can be demonstrated by 
this measurement when $p_{-}\sim 1$ and $p_{+}\sim 0$. When no correlation 
exists between the resonator and the charge qubit, $c_{\downarrow}\sim 0 $
and $p_{\pm }=1/2$.

By measuring the charge qubit, it can also be shown that the states 
$|\uparrow\rangle |-2n\alpha _{0}\rangle $ and 
$|\downarrow\rangle |2n\alpha _{0}\rangle $ are 
in coherent superposition. This measurement starts by applying a 
$\delta $-function $\pi /2$ pulse to the state $\frac{1}{\sqrt{2}} 
(|~\uparrow~\rangle |~-2n\alpha _{0}~\rangle +|~\downarrow~\rangle 
|~2n\alpha_{0}~\rangle ) $, followed by the pulses in Eq.~(\ref{ep_perp}) 
for $n$ time. The state becomes $\frac{1}{2\sqrt{2}}(|+ \rangle 
|\psi_{+}\rangle + |-\rangle |\psi_{-}\rangle )$ with $|\psi_{+}\rangle = 
|-4n\alpha _{0}\rangle + 2 |0\rangle - |4n\alpha _{0}\rangle $ and 
$|\psi_{-}\rangle = |-4n\alpha _{0}\rangle +|4n\alpha _{0}\rangle$. The 
probabilities of the states $|\pm\rangle$ are hence $p_{+}=3/4$ and 
$p_{-}=1/4$. Without the coherence, i.e. that of a mixed state of 
$|\uparrow\rangle |-2n\alpha _{0}\rangle $ and 
$|\downarrow\rangle |2n\alpha _{0}\rangle $, 
the probabilities are $p_{\pm}=1/2$ respectively. Hence measurement of 
the $\sigma_x$ operator probes the coherence of the system.

\begin{figure}[tbh]
\includegraphics[width=7cm,clip]{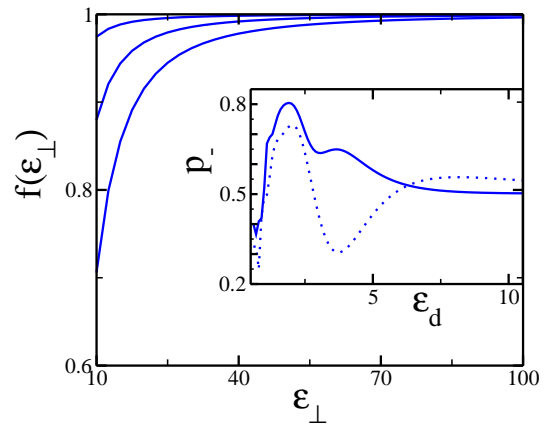}
\caption{The main plot: the fidelity of the amplification versus $\epsilon 
_\perp$ for $n=4,\,8,\,12$ pulses from top to bottom. Inset: the 
probability $p_{-}$ versus $\epsilon_{d}$ at $\epsilon_{z}=4.0\omega_{0}$
(solid line) and $\epsilon_{z}=3.2\omega_{0}$ (dotted line). Here $\lambda_{0}=
20\,\mathrm{GHz}$.}
\label{figure2}
\end{figure}
Ideal situations are assumed in the above discussions with well separated 
energy scales: $\epsilon_\perp \gg \omega_0$ during the amplification and 
$\epsilon_z, \epsilon_d, 8n\alpha_0\lambda_0 \gg \omega_{0}$ during the 
detection. In practice, the frequency of the resonator is around 
$100\,\mathrm{MHz}$; while the amplitudes of the pulses  
$\epsilon_{\perp}$, $\epsilon_{d}$ are upper bounded by the Josephson 
energy $E_{J0}$ of the charge qubit which is typically below
$20\,\mathrm{GHz}$. The dynamics of the resonator may have important effects 
on our scheme. Below, we numerically simulate the dynamics of the 
amplification and detection with the above parameters and in the coordinate 
space of the resonator.

Let the wave function at time $t$ be $|\psi(t)\rangle = \sum _{s}\int dx 
|s\rangle |x\rangle \varphi_s (x,t)$ where $s$ is the state of the charge qubit and 
$\varphi_s (x,t)$ is the wave function of the resonator, with the initial 
state discussed above.
We calculate the fidelity of the amplification process: $f(\epsilon _{\perp 
})=\left| \left\langle \psi _{id}\left( t\right) |\psi \left( t\right) 
\right\rangle \right| ^{2}$, where $|\psi_{id}\rangle$ is the ideal wave 
function generated by the pulses in Eq.~(\ref{ep_perp}). In 
Fig.\ref{figure2}, the fidelity is plotted versus the amplitude of the 
pulse $\epsilon_\perp$. It can be seen that at $\epsilon_{\perp} = 
10\omega_{0}$, the fidelity can be very low with $f=0.7$ after $n=12$ 
pulses; but the fidelity increases rapidly with increasing $\epsilon 
_{\perp}$. At $\epsilon_{\perp}=60\,\omega_{0}$, corresponding to 
$\epsilon_{\perp}=6\,\mathrm{GHz}$, the 
fidelity is $f>0.99$ after $n=12$ pulses. In the detection process, the 
dynamics of the resonator affects the probability $p_{-}$. We simulate the 
detection process at various static bias $\epsilon_{z}$ and with
$\epsilon_{d}$ in the range of $0.5\,\omega_{0}\,-\, 10.5\,\omega_{0}$. Here $8n 
\alpha_{0} \lambda_{0}\approx 1.9\,\omega_{0}$ after $n=12$ pulses. In the 
inset of Fig.\ref{figure2}, $p_{-}$ is plotted versus $\epsilon _{d}$ at 
$\epsilon _{z}=3.2\,\omega_{0}$ and $\epsilon_{z}=4.0\,\omega_{0}$. When 
$\epsilon_{d}\gg 8n\alpha_{0} \lambda_{0}$, the charge qubit flips for both 
the states $|\pm 2n\alpha_{0}\rangle $ and $p_{-}\approx 0.5$. When 
$\epsilon _{d}\sim 8n\alpha_{0}\lambda_{0}$, the off resonance strongly 
affects $p_{-}$. For $\epsilon_{z}=4.0\,\omega_{0}$, a maxumum of $p_{-}$ 
appears at $\epsilon_{d}=1.9\omega _{0}$ with $p_{-}=0.80$, very different 
from the probability without the correlation. For $\epsilon_{z}=3.2\,\omega_{0}$, $p_{-}$ 
oscillates with $\epsilon_{d}$ due to the evolution of the resonator. This 
shows that the correlation between the resonator and the charge qubit can 
be detected even at high resonator frequency.

With a coupling of $20\,\mathrm{MHz}$\cite{Resonator_cooling}, $n\ge 10$ 
flips are required to have $2n\alpha_{0} > 1$, a duration around 
$50\, \mathrm{nsec}$. It is crucial to have a decoherence time longer than 
this duration to successfully generate the entanglement. During the 
amplification process, the charge qubit operates at the degenerate point where 
the charge noise, dominated by the low frequency charge fluctuations, 
causes a decoherence time of microseconds\cite{optimal_point}. In the 
rotating frame of Eq.~(\ref{Htr}), this can be explained by a spectral 
shift: the noise spectrum in this frame is $S^{0}\left( \omega \pm 
E_{J0}/\hbar\right)$ with a shift of $E_{J0}$ from the noise spectrum in the 
lab frame $S^{0}\left(\omega \right)$. The low frequency noise is hence 
screened by the Josephson energy. The mechanical noise of the resonator can 
be described by the quality factor $Q$. At temperature $T=20\,\mathrm{mK}$ 
with $Q=10^{4}$, the dissipation rate is $k_{B}T/Q=50\,\mathrm{KHz}$. The 
decoherence rate is $\tau _{dec}^{-1}\approx (2n\alpha_{0})^{2} 
k_{B}T/Q$, which at $2n\alpha _{0}=5$ gives $\tau _{dec}^{-1} \sim 
1\,\mathrm{MHz}$ and limits the amplification process. Meanwhile, in a 
situation where the phase coherence between the states 
$|\uparrow\rangle |-2n\alpha_{0}\rangle$ and $|\downarrow\rangle |2n\alpha_{0}\rangle$
is not required as in the spin detection\cite{single_spin_detection}, 
the amplitude of the generated coherent states can be bounded by 
the quality factor. Assume after $n_{s} $ 
flippings the amplification saturates. This means that starting from the 
coherent state $|-2 n_{s}\alpha _{0} \rangle$ with the charge qubit at  
$|\uparrow \rangle$, after a time of $\pi/\omega _{0}$, the coherent state 
is $|2n_{s}\alpha _{0} \rangle $. The initial elastic energy is $\hbar 
\omega _{0}\alpha _{0}^{2}\left( 2n_{s}+1\right) ^{2}$, while the final 
elastic energy is $\hbar \omega _{0}\alpha _{0}^{2}\left( 2n_{s}-1\right) 
^{2}$ with an energy loss of $\delta E=8n_{s}\hbar \omega _{0}\alpha 
_{0}^{2}$. The loss is caused by the dissipation: $\delta E=4\pi n_{s}^{2} 
\hbar \omega _{0}\alpha _{0}^{2}/Q$, from which the saturation limit: 
$n_{s}=2Q/\pi $ can be derived.

This scheme involves a generic model of one harmonic oscillator and one 
quantum two level system (spin) coupling via linear interaction 
$\hat{x}\sigma _{z}$, and hence can be generalized to other spin-oscillator 
systems. One example is the single spin detection by magnetic resonance 
force microscopy (MRFM)\cite{single_spin_detection} where the spins near a 
surface interact with the magnetic particle attached to a cantilever and 
affect the vibration of the cantilever. By observing the frequency or
amplitude of the vibration with optical interferometry, the distribution of 
the spins can be detected. In experiments, the resolution of MRFM has been 
improved towards the single spin level\cite{single_spin_detection}. In the 
conventional scheme, the cyclic adiabatic inversion (CAI) 
method\cite{single_spin_detection} is applied where the spins are driven by 
continuous microwave with periodic modulation of the phase of the 
microwave. Our scheme provides an alternative to this approach. By applying 
parametric pulses to flip the spins every half period of the cantilever, 
coherent states of the cantilever with large amplitude can be generated 
within a short time even at weak coupling. Using the same notations as that 
in Eq.~(\ref{Htr}) and assuming a total local spin of $m/2$ near the tip,
after $n$ spin flips in Eq.~(\ref{ep_perp}), the coherent states are $|\pm 
2n m\alpha_{0}\rangle$. This shows a resolution of $\delta \alpha / \delta 
m = 2 n \alpha_{0}$. When $2 n \alpha_{0}\gg 1$ and $n_{s}\ge n$, single spin 
resolution can be achieved. This requires $Q> \pi/4\alpha_{0}$ of the cantilever.

We studied a scheme of generating and detecting Schr\"{o}dinger cat state 
in the coupled resonator and SCPB system. Compared with previous 
works\cite{resonator_SCPB1,resonator_SCPB2}, large amplitude coherent states 
can be obtained at much smaller coupling than the energy of the resonator. 
The scheme provides a practical way of investigating the quantum properties 
of the nanomechanical resonators.

Acknowledgments: We thank I. Wilson-Rae, A. Shnirman and P. Zoller for 
helpful discussions. This work is supported by the Austrian Science 
Foundation, the CFN of the DFG, the Institute for Quantum Information,
and the EU IST Project SQUBIT.


\begin{thebibliography}{99}
\bibitem{ResonatorSET} R.G. Knobel and A.N. Cleland, Nature \textbf{424},
291 (2003); M.D. LaHaye \textit{et al.}, Science \textbf{304}, 74 (2004).

\bibitem{NanoResonator} A.N. Cleland and M.L. Roukes, Nature \textbf{392},
160 (1998); H.G. Craighead, Science \textbf{290}, 1532 (2000); X. Ming
\textit{et al.}, Nature \textbf{421}, 496 (2003).

\bibitem{forcedetection}
V.B. Braginsky and F.Y. Khalili, {\em Quantum Measurement}, Cambridge 
Univ. Press, Cambridge (1992).

\bibitem{Munro_measurement}
W.J. Munro et al, Phys. Rev. A \textbf{66}, 023819 (2002).

\bibitem{QIP_resonator} A.N. Cleland and M.R. Geller, Phys. Rev. Lett.
\textbf{93}, 070501 (2004).

\bibitem{resonator_SCPB1} A. D. Armour, M. P. Blencowe, and K. C. 
Schwab, Phys. Rev. Lett. \textbf{88}, 148301 (2002).

\bibitem{resonator_SCPB2} E.K. Irish and K. Schwab, Phys. Rev. B 
\textbf{68}, 155311 (2003).

\bibitem{feedback_cooling} A. Hopkins \textit{et al.}, Phys. Rev. B. \textbf{%
68}, 235328 (2003).

\bibitem{wilsonrae_cooling}
I. Wilson-Rae, P. Zoller and A. Imamo$\mathrm{\bar{g}}$lu,
Phys. Rev. Lett. \textbf{92}, 075507 (2004).

\bibitem{Resonator_cooling} I. Martin \textit{et al.},
Phys. Rev. B \textbf{69}, 125339 (2004).

\bibitem{IonTrap_Rev} D.J. Wineland \textit{et al.}, J. Res. Natl.   
Inst. Stand. Technol. \textbf{103}, 259 (1998).


\bibitem{charge_qubit_mss} Y. Makhlin, G. Sch\"{o}n, and A. Shnirman,
Rev. Mod. Phys. \textbf{73}, 357 (2001).

\bibitem{qc_braunstein}
S. Lloyd and S.L. Braunstein, Phys. Rev. Lett. \textbf{82}, 1784 (1999).

\bibitem{single_spin_detection} D. Rugar \textit{et al.}, Nature  
\textbf{430}, 329 (2004); J.A. Sidles \textit{et al.}, Rev. Mod. Phys. 
\textbf{67}, 249 (1995).

\bibitem{optimal_point} D. Vion \textit{et al.}, Science \textbf{296}, 886
(2002).
 
\bibitem{kraus_pra}
B. Kraus \textit{et al.}, Phys. Rev. A \textbf{67}, 042314 (2003).
\end{thebibliography}
\end{document}